\documentclass[twocolumn,floatfix,prb]{revtex4}%
\usepackage{mathrsfs,bm,amsmath}
\usepackage{graphicx}

\def\s{{\sigma}}
\def\e{{\epsilon}}
\def\k{{ {\bm k} }}
\def\p{{ {\bm p} }}
\def\q{{ {\bm q} }}
\def\Q{{ {\bm Q} }}
\def\0{{ {\bm 0} }}

\def\w{{\omega}}
\def\a{{\alpha}}
\def\b{{\beta}}

\allowdisplaybreaks[4]

\makeindex

\begin{document}

\title{
$d$- and $p$-wave quantum liquid crystal orders
in cuprate superconductors, $\kappa$-(BEDT-TTF)$_2$X, and
coupled chain Hubbard models:
functional-renormalization-group analysis
}

\author{
Rina Tazai$^{1}$,
Youichi Yamakawa$^{1}$,
Masahisa Tsuchiizu$^{2}$, and
Hiroshi Kontani$^{1}$
}

\date{\today}

\begin{abstract}
Unconventional symmetry breaking without spin order,
such as the rotational symmetry breaking (=nematic or smectic) orders
as well as the spontaneous loop-current orders, 
have been recently reported 
in cuprate superconductors and their related materials.
They are theoretically represented by 
non-$A_{1g}$ symmetry breaking in self-energy,
which we call the form factor $f_{\k,\q}$.
In this paper, we analyze typical Hubbard models
by applying the renormalization-group (RG) method,
and find that various unconventional ordering emerges
due to the quantum interference among spin fluctuations.
Due to this mechanism,
nematic ($\q={\bm0}$) and smectic ($\q\ne{\bm0}$) 
bond orders with $d$-wave form factor 
$f_{\k,\q}\propto\cos k_x-\cos k_y$
appear in both cuprates and $\kappa$-(BEDT-TTF)$_2$X.
The derived bond orders naturally explain the 
pseudogap behaviors in these compounds.
The quantum interference also induces various
current orders with odd-parity form factor.
For example, we find the emergence of the 
charge and spin loop-current orders with $p$-wave form factor
in geometrically frustrated Hubbard models.
Thus, 
rich quantum phase transitions with $d$- and $p$-wave form factors
are driven by the paramagnon interference 
in many low-dimensional Hubbard models.

\end{abstract}

\address{
$^1$Department of Physics, Nagoya University, Furo-cho, Nagoya 464-8602, Japan. \\
$^2$Department of Physics, Nara Women's University, Nara 630-8506, Japan.
}

\maketitle

\section{Introduction}
\label{sec:1}
Various exotic symmetry-breaking,
such as violations of the rotational, time-reversal and inversion symmetries,
have been discovered in many strongly correlated metals,
thanks to the recent progress of experiments.
For example, electronic nematic states 
(=rotational symmetry breaking) without magnetization 
commonly emerge in Fe-based and cuprate superconductors.
These discovered exotic symmetry breaking
are generally called the ``quantum liquid crystal states'', and they
are totally different from conventional local spin/charge density waves
(SDW/CDW) studied so far.
These exotic orders are ``hidden'' 
due to the difficulties in experimental detection,
while they are fundamental states of metals because their
transition temperatures are frequently higher than conventional 
SDW/CDW orders.
In this article, we investigate the rich variety of exotic orderings
in terms of the non-$A_{1g}$ symmetry breaking in self-energy, 
which is represented as the form factor $f_{\k,\q}$,
in a unified way.

\begin{figure}[htb]
\includegraphics[width=8.5cm]{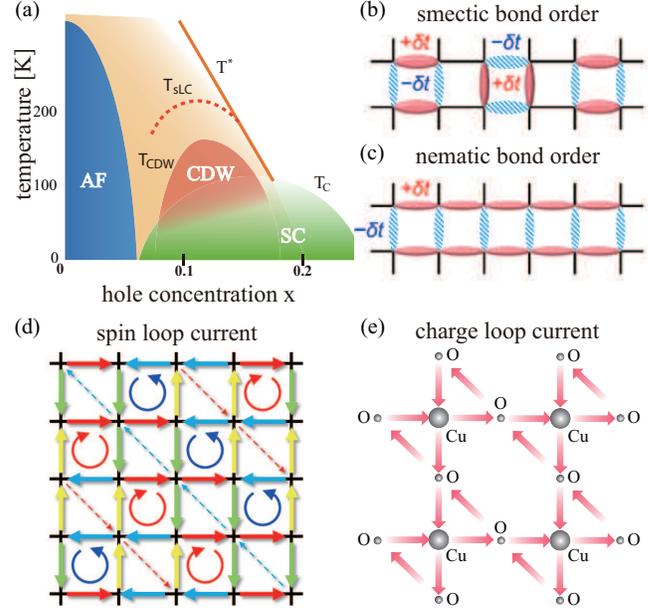}
\caption{(a) Schematic phase diagram of hole-doped cuprates.
(b) Smectic $d$-symmetry bond order at $\q=(\pi/2,0)$ for $T<T_{\rm CDW}$ and
(c) nematic one at $\q={\bm0}$ for $T<T^*$.
(d) Spin loop-current order pattern at $\q=(\pi/2,\pi/2)$.
(e) Intra-unit-cell charge loop-current proposed in Ref. \onlinecite{Varma}.
}
\label{fig1-1}
\end{figure}

Figure \ref{fig1-1} (a) shows a schematic phase diagram of 
cuprate superconductors.
Below $T_{\rm CDW}\sim 200$K,
smectic $p$-orbital charge-density-wave ($p$O-CDW) emerges 
at finite wavevector $\q\approx(\pi/2,0)$ in many compounds
\cite{Y-Xray1,Bi-Xray1,STM-Kohsaka,STM-Fujita}.
The discovery of this smectic $p$O-CDW 
significant progress in theoretical studies.
A natural candidate of the order parameter behind the $p$O-CDW 
is the $d$-symmetry bond order (BO) shown in Fig. \ \ref{fig1-1}(b),
where $\delta t$ represents the modulation of hopping integrals.
Various spin-fluctuation-driven BO mechanisms 
have been proposed 
\cite{Davis:2013ce,Metlitski:2010gf,Husemann:2012eb,Efetov:2013ib,Sachdev:2013bo,Mishra:2015fb,Yamakawa-CDW,Orth:2017,Yamase}.
Also, the pair-density-wave scenarios have been proposed in Refs.
\onlinecite{Berg:2009gt,Fradkin:2015co,Wang:2015iq,Lee:2014ka,Agterberg}.

Another important unsolved issue is the 
origin of the pseudogap in the density-of-states (DoS) below $T^*$.
At present, it is an open problem
whether the pseudogap is a distinct phase or a continuous crossover.
As for the latter case,
short-range spin fluctuations at $T\sim T^*$
can induce the pseudogap due to large quasiparticle damping 
\cite{TPSC,Kotliar,Maier}.
As for the former case,
experimental evidence of the phase transition at
$T^*$ has been accumulated
\cite{Shekhter:2013eh,ARPES-Science2011,Fujimori-nematic,Y-Sato,Hg-Murayama,Shibauchi-nematic}, {\it e.g.},
the ARPES \cite{ARPES-Science2011,Fujimori-nematic}, 
magnetic torque \cite{Y-Sato,Hg-Murayama},
polarized neutron diffraction (PND) \cite{TRSB-neutron1,TRSB-neutron2},
nematic susceptibility \cite{Shibauchi-nematic}
measurements.

The presence or absence of the time-reversal symmetry (TRS) 
in the pseudogap phase has been unsolved for years.
We first discuss candidates of 
TRS preserving order parameter at $T^*$:
Considering the $C_4$ symmetry breaking below $T^*$ \cite{Y-Sato}
and the enhancement of the nematic susceptibility above $T^*$
 \cite{Ishida-nematic},
the $d$-wave nematic ($\q={\bm0}$) order in Fig. \ \ref{fig1-1} (c)
would be naturally expected.
A similar nematic transition is realized 
in many Fe-based superconductors 
\cite{Onari:2012jb,YYamakawa-PRX2016,Onari-FeSe,Onari-B2g,Onari-AFBO}.
On the other hand, pseudogap is not induced by intra-unit-cell orders.
Another candidate of TRS preserving order is the staggered 
($\q={\pi/2,\pi/2}$)
spontaneous spin loop-current (sLC) order shown in Fig. \ \ref{fig1-1}(d).
The sLC order is ``hidden'' in that neither internal magnetic field nor 
charge density modulation is induced, whereas the predicted sLC 
with finite wavenumber naturally gives the Fermi arc structure
and the pseudogap in the DoS.

We also discuss candidates of the TRS broken order parameter at $T^*$:
Figure \ref{fig1-1} (e) depicts the intra-unit-cell ($\q={\bm0}$)
charge loop current (cLC) order 
that accompanies the magnetic field proposed by Varma \cite{Varma}.
Recently, a number of experimental reports for the 
cLC order have been accumulated.
For instance, in quasi 1D two-leg ladder cuprates,  
the PND reveal the broken time-reversal symmetry
\cite{cLC-2leg} and conclude that the cLC appears.
The cLCs in the spin disordered phase are also reported in cuprates \cite{TRSB-neutron1,TRSB-neutron2} 
and iridates \cite{TRSB-iridate}, and their
existence is also supported by the optical second harmonic generation (SHG) 
\cite{SHG-cuprate,SHG-iridate} and Kerr effect \cite{Kerr-cuprate} measurements.

Theoretically, quantum phase transitions in metals 
depicted in Figs. \ref{fig1-1} (b)-(e) are given by the form factor $\delta t_{i,j}^\s$,
which corresponds to spontaneous symmetry breaking in self-energy.
Here, $i$, $j$ represent the sites and $\s$ is the spin index.
Hereafter, we focus on the exotic nature of the ``non-local form factor $i\ne j$''. 
The original hopping integral between sites $i$ and $j$
 is modified to $t_{i,j}+\delta t_{i,j}^\s$.
The Hermite condition leads to the relation
$\delta t_{i,j}^\s = (\delta t_{j,i}^\s)^*$.
Here, we set 
$\delta t_{i,j}^{c(s)}\equiv (\delta t_{i,j}^\uparrow+(-)\delta t_{i,j}^\downarrow)/2$.

\begin{figure}[htb]
\includegraphics[width=9cm]{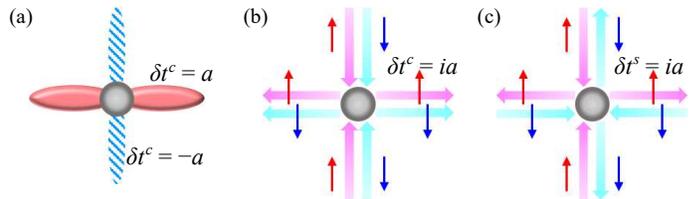}
\caption{Form factor $\delta t_{i,j}^{c,s}$ in each nonlocal order:
(a) BO ($\delta t_{i,j}^{c}=\delta t_{j,i}^{c}=a$), 
(b) cLC ($\delta t_{i,j}^{c}=-\delta t_{j,i}^{c}=ia$), 
and 
(c) sLC ($\delta t_{i,j}^{s}=-\delta t_{j,i}^{s}=ia$). 
Here, $a$ is a real quantity.
}
\label{fig1-2}
\end{figure}

For example, 
the BO is given by a real and even-parity $\delta t_{i,j}^c$,
which is shown in Fig. \ref{fig1-2} (a)
\cite{Bulut,Chubukov,Sachdev,Metzner,Davis:2013ce,Berg:2009gt,Yamakawa-CDW,Tsuchiizu-CDW,Kawaguchi-CDW}.
In contrast, the cLC order is given by a pure imaginary and odd-parity;
$\delta t_{i,j}^c=-\delta t_{j,i}^c = {\rm imaginary}$
\cite{Varma,Affleck,FCZhang,Schultz}.
In this case, $\delta t_{i,j}^c$ represents the fictitious Peierls phase,
and the spontaneous cLC is induced as shown in Figs. \ref{fig1-2} (b).
The cLC order causes the real magnetic field.
In contrast, spin current flows if pure imaginary order parameter
is odd under space and spin inversions;
$\delta t_{i,j}^{s}=-\delta t_{j,i}^{s}= {\rm imaginary}$.
Then, $\delta t_{i,j}^s$ represents the spin-dependent 
fictitious Peierls phase, and therefore spontaneous sLC order 
in Figs. \ref{fig1-2} (c) is induced
\cite{Schultz,Nersesyan,Ozaki,Ikeda,Fujimoto,Sr2IrO4}.

From the microscopic viewpoint, however,
the mechanism of these exotic nonlocal orders is highly nontrivial, 
since the local ($i=j$) SDW/CDW occurs within the mean-field approximation (MFA).
For example, if the Coulomb interaction is local,
the induced order is always local within the MFA.
One may consider that non-local form factors are realized 
in extended Hubbard models with non-local Coulomb interaction.
However, within the MFA, the realization conditions 
of nonlocal orders are severe even in extended $U$-$V$-$J$ Hubbard model
 \cite{Nersesyan}.
These facts indicate the importance of non-local effective 
interaction due to beyond-mean-field many-body effects,
called the vertex corrections (VCs).
This is the main issue of the present article.

\begin{figure}[htb]
\includegraphics[width=4.5cm]{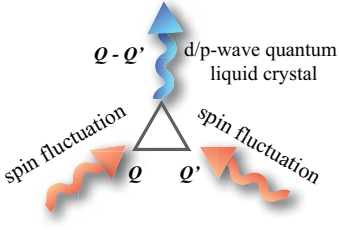}
\caption{Interference mechanism due to spin fluctuations. It causes
exotic $d$- an $p-$wave quantum liquid crystal phases such as 
nematic/smectic BO, sLC, and cLC orders.}
\label{fig-interference}
\end{figure}

Recently, important roles of the VCs on the nonlocal orders
have been revealed step by step.
The nematic order in Fe-based superconductors
is induced by the Aslamazov-Larkin (AL) VCs
\cite{Onari:2012jb,YYamakawa-PRX2016,Onari-FeSe,Onari-B2g,Onari-AFBO},
which are significant near the magnetic quantum-critical-point (QCP).
The physical meaning of the AL-VCs is the ``quantum interference'' 
between different spin fluctuations at ${\bm Q}$ and ${\bm Q}'$,
which is depicted in Fig. \ref{fig-interference}.
Due to this mechanism,
non-local order at ${\bm Q}-{\bm Q}'$ is established.
This mechanism has been applied to many strongly correlated metals
to explain various hidden orders
\cite{Onari:2012jb,YYamakawa-PRX2016,Onari-FeSe,Onari-B2g,Onari-AFBO}.
In cuprate superconductors,
emergence of the BO and the sLC order 
has been discussed based on the paramagnon-interference (AL-VC) mechanism
\cite{Yamakawa-CDW,Kawaguchi-CDW,Tsuchiizu-CDW,Tsuchiizu:2016ix}.
In addition, other spin-fluctuation-driven mechanisms have been 
successfully applied
\cite{Davis:2013ce,Metlitski:2010gf,Husemann:2012eb,Efetov:2013ib,Sachdev:2013bo,Mishra:2015fb,Orth:2017}.

The renormalization group (RG) theory
is a very powerful method to study the quantum interference, 
because huge numbers of parquet-type diagrams
are generated by solving the RG differential equation. 
Although conventional $N$-patch RG 
\cite{Zanchi:1996ul,Zanchi:1998ua,Zanchi:2000ua} is applicable for a model 
with simple band dispersion, this constraint is alleviated
by combining the RG and the constrained randon-phase-approximation (cRPA).
Using this RG+cRPA method, we can calculate
the general charge (spin) susceptibilities
with non-local form factor, $\chi^{c(s)}_f(\q)$,
by including higher-order VCs.
The realized order with form factor $f$  
at wavevector $\q$ is determined under the condition of
maximizing the function $\chi^{c(s)}_f(\q)$.


In this article,
we perform the RG analysis of exotic ``quantum liquid crystal states'' 
described by non-$A_{1g}$ and non-local form factor in cuprate superconductors, 
$\kappa$-(BEDT-TTF)$_2$X, and coupled chain Hubbard models.
It is clarified that 
the paramagnon interference mechanism causes 
rich quantum phase transition with $d$-wave and $p$-wave form factors
in typical low-dimensional Hubbard models.
This paper is organized as follows:
In Sect. \ref{sec:2}, we explain the formalism of the RG method,
based on which we derive general charge (spin) channel 
susceptibility with form factor, $\chi_f^{c(s)}(\q)$.
We explain how to derive the optimized form factor $f$.
In Sects. \ref{sec:3} and \ref{sec:4}, we discuss the BO order and sLC order formation 
in cuprate superconductor and $\kappa$-(BEDT-TTF)$_2$X, respectively.
In Sect \ref{sec:5}, we discuss the cLC order induced in 
the quasi 1D Hubbard model with geometrical fluctuation.
The BO, sLC and cLC orders are induced by paramagnon interference 
mechanism, which is totally dropped in the MFA.
The discussions and summary are presented in Sect. \ref{sec:6}.

\section{renormalization group theory}
\label{sec:2}

``How to treat many-body effects'' is one of the long-standing problems in strongly correlated electron systems.
Until now, a number of theoretical approaches have been generated \cite{George-rev,Yokoyama,Varma2,various1,various2}, such as
dynamical mean-field theory (DMFT)  and variational Monte Carlo (VMC) studies.
One fundamental approach is perturbation theory based on diagrammatic expansion.
However, it is hopeless to consider all possible diagrams, which continue to infinite order. 
Then, low-order perturbation theory fails to explain strongly correlated systems, such as the Mott transition in cuprates.
For long time, the random-phase-approximation (RPA),
in which the infinite series of particle-hole loop diagrams are considered,
has been used to explain various magnetic transition.

On the other hand, RPA cannot explain recently discovered exotic non-local transitions, since these phenomena originate from mode coupling via spin-, charge-, and orbital-degrees of freedom in nature. 
Therefore, it is necessary to consider the vertex corrections (VCs) neglected in conventional RPA.

Renormalization group (RG) is a powerful tool to calculate the VCs accurately
 in an unbiased way. 
Thanks to the recent theoretical improvement, exotic non-local susceptibilities $\chi^{c(s)}_f (\q)$ with form factors $f_{\k,\q}$ are analyzed by RG with high accuracy.
For instance, in this article, we calculate the non-local susceptibilities in  cuprates for
nematic/smectic $d$-symmetry BO as well as $p$-symmetry sLC 
and cLC ordered phase.
Hereafter, we explain the optimization of form factors within the RG scheme.

\subsection{RG formalism}
Here, we explain the RG method based on one-orbital Hubbard models
whose interaction part is given by
\begin{eqnarray}
\hat{H}_{I}=&&\sum_{\k_{i},\sigma_i}
\Gamma^{\sigma_1 \sigma_2 \sigma_3 \sigma_4}_{\k_1 \k_2\k _3 \k_4}
c^{\dagger}_{\k_1\sigma_1}
c_{\k_2\sigma_2}
c^{\dagger}_{\k_4\sigma_3}
c_{\k_3\sigma_4},
\label{eq:34}
\end{eqnarray}
where $c^{\dagger}_{\k\sigma}$ is the creation operator of the electron,
and $\hat \Gamma$ is fully antisymmetrized four-point bare vertex.
In the case of on-site Coulomb repulsion $U$,
the non-zero components of $\hat \Gamma$ are given as
$\Gamma^{\sigma \sigma \bar{\sigma} \bar{\sigma}}=
-\Gamma^{\sigma \bar{\sigma} \bar{\sigma} \sigma}=U/4$,
so the requirement by
the Pauli exclusion principle is satisfied.
When the system has SU(2) symmetry in the spin space,
the following relation is satisfied,
\begin{eqnarray}
\Gamma^{\sigma \sigma \sigma \sigma}_{\k_1\k_2\k_3\k_4}-\Gamma^{\sigma \sigma \sigma \sigma}_{\k_1\k_3\k_2\k_4}=\Gamma^{\sigma \sigma  \bar{\sigma} \bar{\sigma}}_{\k_1\k_2\k_3\k_4}-\Gamma^{\sigma  \sigma \bar{\sigma} \bar{\sigma}}_{\k_1\k_3\k_2\k_4}.
\label{eqn:SU2}
\end{eqnarray}
Note that $\Gamma^{\sigma \sigma \sigma \sigma}$ and $\Gamma^{\sigma \sigma \bar{\sigma} \bar{\sigma}}$ stand for four-point vertex function with parallel- ($g^{\parallel}$) and anti-parallel-spin ($g^{\perp}$) in $g$-ology theory \cite{Emery,Bourbonnais,Kishine1,Kishine4,
Suzumura,Suzumura2,SSolyom}.
Thus, Eq. (\ref{eqn:SU2}) is equivalent to the relation $g^{\parallel}_{1}-g^{\parallel}_{2}=g^{\perp}_{1}-g^{\perp}_{2}$, which we will use in the Sect \ref{sec:5}.
Also, tensor $\Gamma$ is uniquely decomposed into spin- and charge-channels as
\begin{eqnarray}
\Gamma_{\k_1\k_2\k_3 \k_4}^{\sigma \sigma' \rho \rho'}
=\frac{1}{2}\Gamma^{s}_{\k_1\k_2\k_3 \k_4}
\vec{\sigma}_{\sigma \sigma'} \cdot \vec{\sigma}_{\rho' \rho}
+\frac{1}{2}\Gamma^{c}_{\k_1\k_2\k_3 \k_4}
\delta_{\sigma \sigma'}\delta_{\rho' \rho},
\label{eqn:Gamma2}
\end{eqnarray}
where $\vec{\sigma}$ is the Pauli matrix vector.
\begin{figure}[htb]
\includegraphics[width=9cm]{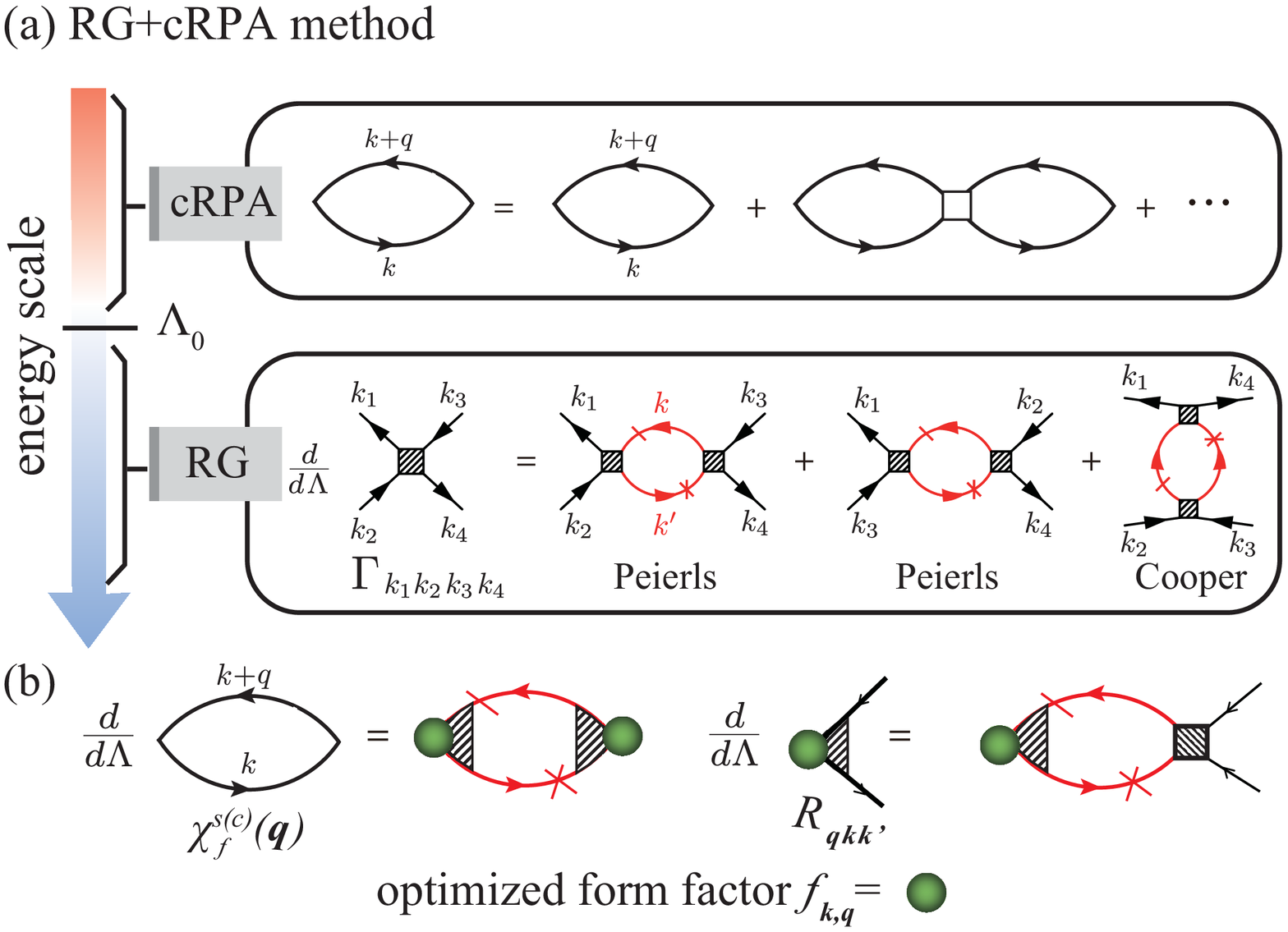}
\caption{(a) Diagrammatic explanation of the RG+cRPA method.
(b) RG equation of the susceptibility $\chi^{c(s)}_f(\q)$ 
and three-point vertex function $R_{\q\k\k'}$ with 
the form factor $f_{\k,\q}$.}
\label{fig:RG}
\end{figure}

The RG concept is based on the adiabatic condition, in which 
the energy scale of viewing the electron system gradually changes.
For this purpose, the logarithmic energy cutoff is introduced as
\begin{eqnarray}
\Lambda=\Lambda_{0}e^{-l} \hspace{10pt} (l\ge0).
\label{eqn:logene}
\end{eqnarray}
Also, the Green function with energy cutoff is defined as
\begin{eqnarray}
G(k)\equiv (i\e_n-\xi_\k)^{-1} \Theta(\Lambda-|\xi_{\k}|),
\end{eqnarray} 
where $\xi_{\k}$ is the energy dispersion of the
electron, and $k=(\k,\e_n)$ with wavevector $\k$ and 
fermion Matsubara frequency $\e_n$.
$\Theta$ is the Heaviside step function reflecting
high energy cutoff within the RG framework.
Based on the path integral RG formula,
fermionic field operators
on the energy-shell $\Lambda -d \Lambda < |\xi_{\k}| < \Lambda$
are integrated out.
Then, the effective low-energy four-point vertex 
with $|\xi_{\k}| < \Lambda- d \Lambda$ is obtained.
This complex procedure is automatically performed by solving the following  differential equation, so-called RG equation.
Within the one-loop approximation, it is given by
\begin{eqnarray}
\hspace{-5pt}\frac{d}{d\Lambda}
\Gamma_{\k_1 \k_2 \k_3 \k_4}
 &=&
-\frac{T}{N}\sum_{\k,\k',\e_n,\e_{m}} 
\frac{d}{d\Lambda}
\left[ G(\k,\e_n) \, G(\k',\e_m) \right] 
\nonumber \\
& &\! \! \! \! \! \! \! \! \! \! \! \! \! \! \! \! \! \!
 \times \Bigl\{ \Bigl[
\Gamma_{\k_1\k_2\k\k'} \Gamma_{\k\k'\k_3\k_4}
-\Gamma_{\k_1\k_3\k\k'} \Gamma_{\k\k'\k_2\k_4}
\Bigr]\delta_{\e_n,\e_m}
\nonumber \\
& &\! \! \! \! \! \! \! \! \! \! \! \! \! \! \! \! \! \!
+ \frac{1}{2} \Gamma_{\k_1\k\k'\k_4}\Gamma_{\k\k_2\k_3\k'} \delta_{\e_n,-\e_{m}}
\Bigr\}.
\label{eqn:S-RG}
\end{eqnarray}
The first and second terms in the right-hand-side originate from Peierls-channel
scattering, while the third one corresponds to the Cooper-channel scattering.
Here, we employ the Wick-ordered scheme \cite{Wick}, in which the 
cutoff function $\Theta_<= \Theta(\Lambda-|\xi_{\k}|)$
is used for the Green function \cite{Metzner}.
Thus, the VCs due to the higher-energy processes 
are included more accurately than the Kadanoff-Wilson scheme
for $\Theta_>= \Theta(|\xi_{\k}|-\Lambda)$
in Ref. \onlinecite{Tsuchiizu:2016ix}.

The renormalization procedure of the four-point vertex function is
summarized as follows; 
(i) At the starting point of $\Lambda=W$ (bandwidth), $\Gamma$ takes $U$.
(ii) $\Gamma$ is gradually renormalized toward $\Lambda \rightarrow 0$
by following the RG equation.
(iii) Finally, we obtain the effective low-energy $\Gamma$ including higher-order many-body effects.
The final $\Gamma$ is essentially equal to the one by solving the
parquet equation introduced by Abrikosov \cite{parquet}.

In general, the best way to consider the many-body processes is applying the RG method 
(i)-(iii) to the full energy region of electron states by putting $\Lambda_0=W$. 
However, in this way, the numerical accuracy
is not ensured since it is impossible to fully consider $\k$- and $\e_n$-dependences of $\Gamma$.
To improve the numerical accuracy, $N$-patch RG has been invented
\cite{Halboth:2000vm,Metzner,Zanchi:1996ul,Zanchi:1998ua,Zanchi:2000ua}, in which the momentum space of the electron system is divided into finite
$N$-patches.  While it is, it fails to explain two-dimensional electron systems, in which $k$-dependence of band structure
plays important roles in nature. 
Therefore, more reliable RG framework is required to study
two-dimensional strongly correlated electron systems such as cuprate
superconductors.

\subsection{RG+cRPA with form factor}
In an effort to establish more reliable RG to apply
two-dimensional electrons, the RG+cRPA method
has been recently developed \cite{Penc:1994uh,Tsuchiizu:2002eg,Tsuchiizu:2004ct,
Tsuchiizu:2013gu,Tsuchiizu:2015cs,Tsuchiizu:2016ix,Tazai-FRG}.
We show the diagrammatic explanation of the RG+cRPA method
in Fig. \ref{fig:RG} (a).
In RG+cRPA method, the energy scale of electron system is divided into 
two different region by putting $\Lambda_0<W$.
The higher energy region with $\Lambda_0<\Lambda<W$ is
considered by RPA with fine $\k$-mesh by dropping the VCs. 
On the other hand, the lower energy region with $\Lambda<\Lambda_0$ is considered by
RG scheme. This hybrid method is based on the intuitive idea that
higher-order many-body effects become significant only in low-energy region.
Thanks to RG+cRPA, the numerical accuracy of the susceptibilities
is drastically improved even in the weak-coupling region.

Here, we consider the important roles of the form factor $f_{\k,\q}$
within the RG+cRPA scheme to explain non-local symmetry breaking 
such as nematic/smectic $d$- and $p$-symmetry orders.
The charge- (spin-) channel static susceptibility
with form factor is given by
\begin{eqnarray}
\chi^{c(s)}_{ff'}(\bm q)
&=&
\frac{1}{2}
\int_0^{\beta} \! d\tau \,
\left\langle 
O_f^{c(s)}(\bm q,\tau)O_{f'}^{c(s)}(-\bm q,0) \right\rangle, \label{eqn:chif} \\
O_{f}^{c(s)}(\bm q)
& \equiv & \sum_\k f_{\k,\q} \left\{ c_{\k_+ \uparrow}^\dagger c_{\k_- \uparrow}
+(-)c_{\k_+ \downarrow}^\dagger c_{\k_- \downarrow}\right\},
\end{eqnarray}
where 
$\k_\pm = \k \pm \q/2$.
We denote $\chi^{c(s)}_{f}\equiv \chi^{c(s)}_{ff}$ hereafter.
The RG equation of the charge- (spin-) channel susceptibilities
with respect to the form factor $f_{\k,\q}$ are given as
\begin{eqnarray}
\frac{d}{d\Lambda}\chi^{c(s)}_f(\q)
&=&\frac{T}{N}\sum_{\k,\k'\e_n}
\frac{d}{d\Lambda}\left[
G(\k,\e_n) G(\k',\e_n) \right] \delta_{\k',\k+\q}  
\nonumber \\
&&\times R^{c(s)}_{f, \q \k \k'} R^{c(s)}_{f, -\q \k' \k},
\label{eq:chi-RG}
\\
\frac{d}{d\Lambda}R^{c(s)}_{f, \q \k \k'}
&=&\frac{T}{N}\sum_{\p,\p',\e_n} 
\frac{d}{d\Lambda}\left[G(\p,\e_n) G(\p',\e_n) \right] \delta_{\p',\p+\q}
\nonumber \\
&&\times R^{c(s)}_{f, \q \p \p'}\Gamma^{c(s)}_{\p \p' \k \k'},
\label{eq:R-RG}
\end{eqnarray}
where $R^{c(s)}_f$ is the three-point vertex function of 
charge- (spin-) channel with form factor $f$,
and its initial function is 
\\
\begin{eqnarray}
R^{c(s)}_{f, \q \k \k'}(l=0)=f_{(\k+\k')/2, \q} 
+ \mbox{[cRPA correction]}.
\label{eq:R0}
\end{eqnarray}
The diagrammatic RG flows of 
$\chi^{c(s)}_f$ and $R^{c(s)}$ are given in  Fig.\ \ref{fig:RG} (b).
Based on the Lagrange multipliers method,
we optimize the form factor $f_{\k, \q}$ so as to maximize the
susceptibility.
For this purpose, we introduce the Fourier expansion of the form factor as
\begin{eqnarray}
f_{\k, \q}=\sum_{n,m=1}^{7}2 a_{n m}^{\q}  h_{n}(k_x) 
h_m(k_y),
\end{eqnarray}
where $h_{n}(k)=\{ \frac{1}{\sqrt{2}}, \cos k, \cos 2k, \cos 3k,
\sin k, \sin 2k, \sin 3k \}$ for $n=1-7$, respectively.
The coefficient $a_{n m}^{\q}$ is optimized under the condition
$\frac{1}{N}\sum_\k|f_{\k,\q}|^2=1$ 
by solving the following eigen equation,
\begin{eqnarray}
\sum_M \, \chi_{LM}^{c(s)}(\q) a^{\q}_{M} =\lambda \,a_{L}^{\q},
\end{eqnarray}
where each index $M \equiv (m,m')$ and $L \equiv (l,l')$ takes $1-7^2$.
Here, $\chi_{LM}^{c(s)}$ is the susceptibility with respect to 
the form factors $f=2h_m(k_x)h_{m'}(k_y)$ and $f'=2h_l(k_x)h_{l'}(k_y)$
in Eq. (\ref{eqn:chif}).
The eigenvalue $\lambda$ corresponds to the 
undetermined multiplier in the Lagrange multipliers method.
In the following sections, we discuss the exotic non-local orders in cuprates
based on the present improved RG+cRPA method with form factor $f_{\k, \q}$.

\section{$d/p$-wave bond/sLC order in cuprates}
\label{sec:3}
\begin{figure}[htb]
\includegraphics[width=9cm]{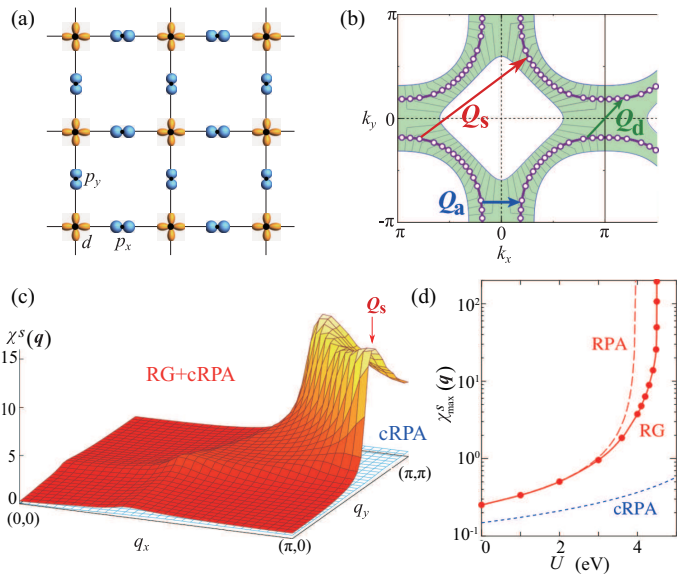}
\caption{
(a) $d$-$p$ Hubbard model.
(b) Fermi surface (FS) of the present YBCO model.
The lower-energy region ($|\xi_{\k}|<\Lambda_0=0.5$ eV)
is denoted by the shaded area.  
The $N$-patch discretization for 
$N =64$ is shown, whereas we set $N = 128$ 
in the present numerical study.
(c) Obtained spin susceptibility $\chi^{s}(\bm q)$.
(d) $U$ dependencies of 
$\chi^{s}_\mathrm{max} [\equiv \chi^s(\bm Q_{rm S})]$ 
given by the RG+cRPA method and by the RPA.  
The initial spin susceptibility given by the cRPA is very small.  
}
\label{cuprate1}
\end{figure}

To understand the origin of exotic phase transitions
in cuprate superconductors exhibited in Fig. \ref{fig1-1},
we perform the RG+cRPA analysis by focusing on the
importance of the quantum interference 
(see Fig. \ref{fig-interference})
\cite{Tsuchiizu-CDW,Tsuchiizu:2016ix}.
We investigate a three-orbital $d$-$p$ Hubbard model
shown in Fig.\ \ref{cuprate1} (a) for YBCO 
\cite{Bulut,Yamakawa-CDW,Thomson:2015ie}.
Its Hamiltonian is given by
\begin{equation}
H_{dp}
=
\sum_{\bm k, \sigma}
\bm c_{\bm k, \sigma}^\dagger \,
\hat h_0(\bm k)  \,
\bm c_{\bm k, \sigma}^{}
+U\sum_{\bm j}
n_{d, \bm j,\uparrow}
n_{d, \bm j,\downarrow},
\end{equation}
where $\bm c_{\bm k, \sigma}^\dagger=
(d_{\bm k,\sigma}^\dagger, p_{x,\bm k,\sigma}^\dagger, 
 p_{y,\bm k,\sigma}^\dagger)$
is the creation operator for the electron 
on $d_{x^2-y^2}$, $p_x$, and $p_y$ orbitals 
with wavevector $\bm k$
  and spin $\sigma$.
$U$ is the Coulomb interaction on $d$-orbital,
and $n_{d,\bm j,\sigma}=d^\dagger_{{\bm j}\sigma} d_{{\bm j}\sigma}$.
In the kinetic term $\hat h_0(\bm k)$,
we introduce the third-nearest $d$-$d$ hopping $-0.1$ eV 
into the first-principles $d$-$p$ model for La$_2$CuO$_4$ 
\ \cite{Hansmann:2014ib}
in order to reproduce YBCO-like Fermi surface (FS)
depicted in Fig.\ \ref{cuprate1} (b)
 \cite{Yamakawa-CDW}. 
The electron filling is set to $n = n_d + n_p = 4.9$, 
corresponding to the hole number $x = 0.1$.

Figure \ \ref{cuprate1} (c) shows the spin susceptibility
$\chi^{s}(\bm q)$ obtained by the RG+cRPA method.
The obtained strong spin fluctuations at
$\bm Q_\mathrm{S} =(\pi-\delta_\mathrm{s},\pi)$ and 
$\bm Q_\mathrm{S}' =(\pi,\pi-\delta_\mathrm{s})$
are consistent with the neutron inelastic scattering measurements.
As increasing $U$, 
$\chi^{s}_\mathrm{max} \equiv \chi^{s}({\bm Q_{\rm S}})$
develops monotonically and diverges 
at $U=U^\mathrm{cr}(\approx 4.5 \mbox{ eV})$,
as shown in Fig.\ \ref{cuprate1} (d).
Thanks to the numerical accuracy of the RG+cRPA method,
$\chi^s_\mathrm{max}$ perfectly follows the RPA result
for a wide weak-coupling region ($U<4$ eV).

As seen from Figs.\ \ref{cuprate1} (c) and (d), 
the cRPA contribution for $\Lambda_0=0.5$ eV in the initial value
is small but very important for the RG analysis
 \cite{Tsuchiizu:2013gu,Tsuchiizu:2015cs}.
We verified in Ref. \onlinecite{Tsuchiizu-CDW} that 
the numerical results by the RG+cRPA method
are \textit{qualitatively} similar to those by conventional patch-RG method, 
whereas the numerical accuracy is well improved.

Next, we investigate the following
$B_{1g}$-symmetry ($d$-symmetry) charge susceptibility 
for $p$-electrons,
\begin{eqnarray}
\chi^{p\mbox{-}\mathrm{orb}}_{d}(\bm q)
&=&
\frac{1}{2}
\int_0^{\beta} \! d\tau \,
\left\langle 
n^{p\mbox{-}\mathrm{orb}}_d(\bm q,\tau)
n^{p\mbox{-}\mathrm{orb}}_d(-\bm q,0)
\right\rangle, \nonumber \\
n^{p\mbox{-}\mathrm{orb}}_d(\bm{q}) &\equiv & n_{x}(\bm{q}) - n_{y}(\bm{q}) 
\end{eqnarray}
where $n_{x(y)} (\bm q ) =  \sum_{\bm k, \sigma}
 p_{x(y),\bm{k}\sigma}^\dagger p_{x(y),\bm{k+q},\sigma}$ is the 
$p$-orbital charge-density-wave ($p$O-CDW) operator.

\begin{figure}[htb]
\includegraphics[width=8cm]{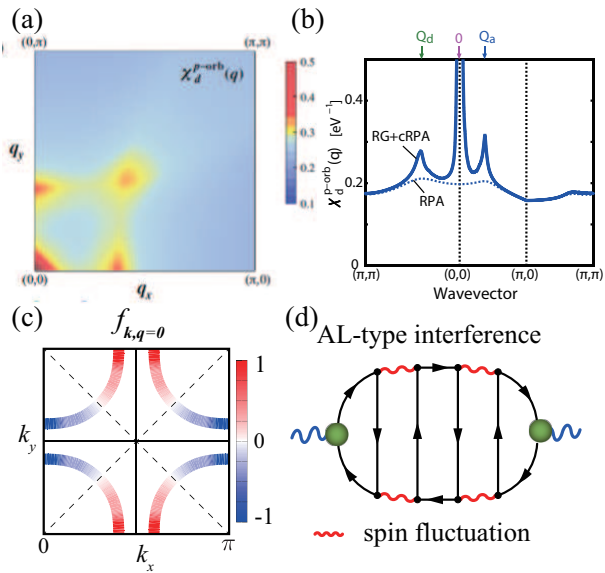}
\caption{
(a) (b) Obtained $p$O-CDW susceptibility 
$\chi^{p\mbox{-}\mathrm{orb}}_d(\bm q)$.
The RPA result is also shown for comparison in (b).
The obtained peak at $\q={\bm Q_\mathrm{a}}$, 
which corresponds to the nesting vector in 
Fig.\ \ref{cuprate1} (b), is consistent with experimental CDW wavevector.  
(c) The optimized form factor $f_{\k,\q=\bm0}$ on the FS, 
which has the $d$-symmetry.
(d) Example of AL-type vertex corrections that give large
$p$O-CDW susceptibility. }
\label{cuprate2}
\end{figure}

Figures\ \ref{cuprate2} (a) and (b) exhibit the obtained
$\chi^{p\mbox{-}\mathrm{orb}}_{d}(\bm q)$
by the RG+cRPA method for $U=4.32$ eV at $T=0.1$ eV.  
The obtained large peaks at 
$\bm q=\bm 0$, $\bm Q_\mathrm{a}$, and $\bm Q_\mathrm{d}$
originate from the VCs, since 
the RPA result is small and non-singular
as seen in Fig.\ \ref{cuprate2} (b).
We see that the highest peak locates at $\bm q=\bm 0$.
This is consistent with the experimental uniform
nematic transition at $T^* \ (>T_{\rm CDW})$
\cite{Y-Sato}.
The second highest peak locates at $\bm q=\bm Q_\mathrm{a}$,
which naturally explains the CDW phase below $T_{\rm CDW}$
in Fig. \ref{fig1-1} (a). 
Note that the temperature $T=0.1$ eV is comparable to $T^*\sim300$ K
if the mass-enhancement factor $m^*/m_{\rm band}\sim 3$ is taken into account.
%

The enhancement of $\chi^{p\mbox{-}\mathrm{orb}}_{d}(\bm q)$ 
in Figs. \ref{cuprate2} (a) and (b)
is also obtained in the single-orbital Hubbard model,
as the enhancement of $d$-wave bond susceptibility.
To explain this fact, 
we investigate the 
$d$-electron charge susceptibility in the $d$-$p$ Hubbard model
with form factor $\chi^c_f(\q)$,
which is introduced in Eq. (\ref{eqn:chif}).
We optimize the form factor by following Sect. \ref{sec:2}.
The numerically optimized $f_{\k,\q}$ at $\q=\bm 0$ 
is shown in Fig.\ \ref{cuprate2}(c), which has the $B_{1g}$-symmetry.
Its Fourier transformation 
gives the modulation of the effective hopping integrals,
called the $d_{x^2-y^2}$-wave BO.
The $\k$-dependence of $f_{\k,\bm0}$ 
in Fig.\ \ref{cuprate2}(c) is similar to that of 
the expectation value of the operator
$n^{p\mbox{-}\mathrm{orb}}_d({\bm0})$ on the FS that is proportional to 
$|u_x(\k)|^2-|u_y(\k)|^2$.
Here, $|u_{x(y)}(\k)|^2$ is the weight of $p_{x(y)}$ orbital
at the Fermi momentum $\k$.
Thus, the $p$O-CDW obtained in the $d$-$p$ Hubbard model
is essentially equivalent to the $d$-wave BO
in the single-orbital Hubbard model.

Next, 
we discuss the physical picture of the origin of BO 
obtained by the RG study.
To find out the significant quantum process,
it is useful to perform the diagrammatic calculations
and to compare the obtained results with the RG results.
For this purpose,
we develop the density-wave (DW) equation method
\cite{Onari-FeSe,Kawaguchi-CDW,Onari-B2g,Onari-AFBO}.
Using this method, we can obtain the most divergent susceptibility
with the optimized form factor, by including higher-order VCs.
It is clarified that Aslamazov-Larkin (AL) type VCs
shown in Fig.\ \ref{cuprate2} (d) are the origin of
the enhancement of the $p$O-CDW susceptibility.
Here, red wave lines represent the dynamical spin susceptibility,
and the paramagnon interference 
given by the convolution of $\chi^s$'s,
$C_\q\equiv \sum_\p\chi^s(\p+\q)\chi^s(\p)$,
becomes large 
$\q\approx \Q_{\rm S}-\Q_{\rm S}= {\bm 0}$ and 
$\q\approx \Q_{\rm S}-\Q_{\rm S}'\approx \Q_\mathrm{d}$.
This quantum interference gives rise to the BO formation,
see Fig. \ref{fig-interference}.
(Note that moderate peak at $\bm Q_\mathrm{d}$ is caused by the 
single-fluctuation-exchange processes called the Maki-Thompson (MT) VC
 \cite{Sachdev:2013bo,Mishra:2015fb}.)

Both the RG+cRPA method and the DW equation method
conclude the emergence of the $d$-wave BOs at 
$\q={\bm0}$ and $\Q_\mathrm{a}$
in several single-orbital Hubbard models.
Thus, the nematic ($\q={\bm0}$) and smectic ($\q\ne{\bm0}$) 
BO formations due to paramagnon interference are expected to be general 
in many strongly correlated metals.
The DW equation method was originally developed 
to explain the electronic nematic order in Fe-based superconductors
\cite{Onari:2012jb,YYamakawa-PRX2016,Kontani:2014ws}.


It is noteworthy that 
the DW equation method also predicts the emergence of
the sLC order described by the spin-channel form factor
\cite{HKontani-sLC}.
The predicted transition temperature $T_{\rm sLC}$
is higher than $T_{\rm CDW}$.
The obtained sLC in real space is shown in Fig. \ref{fig1-1} (d).
The sLC order is ``hidden'' in that neither internal magnetic field nor 
charge density modulation is induced, whereas the predicted sLC 
naturally explains the pseudogap in the DoS at $T^*$.
It is an important future issue to study the 
general spin-channel susceptibility with nonlocal form factor
based on the RG+cRPA method.

\section{$d$-wave bond order in $\kappa$-(BEDT-TTF)$_2$X}
\label{sec:4}
\begin{figure}[htb]
\includegraphics[width=8cm]{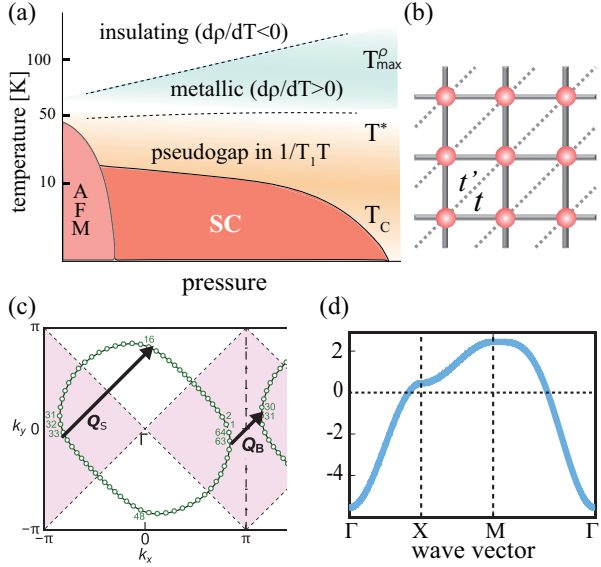}
\caption{
(a) Schematic $P$-$T$ phase diagram of $\kappa$-(BEDT-TTF)$_2$X
\cite{Raman}.
(b) Anisotropic triangular dimer Hubbard model.
(c) FS and (d) band structure of the dimer Hubbard model
at half filling with $t'/t=0.5$.
}
\label{fig-BEDT1}
\end{figure}

The layered organic superconductor $\kappa$-(BEDT-TTF)$_2$X
has been attracting great attention
as a similar substance to cuprate superconductors.
Schematic $P$-$T$ phase diagram is depicted in 
Fig. \ref{fig-BEDT1} (a) \cite{Raman}.
Under pressure, unconventional superconductivity ($T_{\rm c}\gtrsim10$K) 
appears next to the antiferro magnetic (AFM) phase \cite{Kanoda-rev,Kanoda-rev2}.
(In X=Cu[N(CN)$_2$]Br and X=Cu(NCS)$_2$,
metallicity and superconductivity appear even at ambient pressure.)
$T^\rho_{\rm max}$ is the metal-insulator crossover temperature
observed in the resistivity.
$T^*$ is the pseudogap temperature, below which 
the NMR relaxation ratio $1/T_1T$ 
\cite{Kanoda-rev,Kanoda-rev2}
and the DoS measured by the STM
\cite{Nomura-STM}
exhibit the gap-like behaviors.
The nature of the pseudogap and its relation to the superconductivity
has been a central mystery in the unconventional metallic states 
in $\kappa$-(BEDT-TTF)$_2$X.

To study the origin of the pseudogap,
we introduce the anisotropic triangular lattice dimer Hubbard model
shown in Fig. \ref{fig-BEDT1} (b).
Each site in the dimer model is composed of the 
anti-bonding molecular orbital of the BEDT-TTF molecule dimer
This is the simplest effective model 
for $\kappa$-(BEDT-TTF)$_2$X; $\hat{H}=\hat{H}_{0}+\hat{H}_{I}$
\cite{Kino-Fukuyama}.
The kinetic term is given by 
$\hat{H}_{0}=\sum_{\k\sigma}\xi_{\k}c^{\dagger}_{\k\sigma}c_{\k\sigma}$
with $\xi_{\k}=2t(\cos k_x+\cos k_y)+2t'\cos(k_x+k_y)$.
Here, we set the hopping integrals in Fig. \ref{fig-BEDT1} (b)
as $(t,t')=(-1,-0.5)$.
We verified that similar numerical results are obtained for 
$t'/t=0.5\sim 0.8$,
which is realized in many $\kappa$-(BEDT-TTF) families
\cite{McKenzie}.

In this dimer Hubbard model,
both RPA and FLEX approximation predict the emergence of 
spin fluctuations at $\Q_{\rm S}\approx(\pi ,\pi)$, consistently with 
experimental staggered AFM order 
\cite{Schmalian-ET,Kino-ET,Kondo-ET,Kontani-ET}.
These spin fluctuations mediate the
$d_{x^2-y^2}$-wave superconductivity
\cite{Schmalian-ET,Kino-ET,Kondo-ET,Kontani-ET}.
(Based on more realistic four-site Hubbard models,
the $d_{xy}$-wave state can be obtained in case of
weak spin fluctuations at $\q\sim(\pi,0)$
\cite{Kuroki-ET}.)

In the following numerical study,
we set the energy unit $|t|=1$, 
and put the temperature $T=0.05$
and the electron filling $n=1$ ($\mu=0.55$).
The FS and the band structure 
are presented in Figs. \ref{fig-BEDT1} (c) and (d), respectively.
The patch indices ($1\sim64$) are shown on the ellipsoid electron pockets.
The total band width is $W\sim 10$ (in unit $|t|=1$), and
$|t|$ corresponds to 0.05eV since $W\sim 0.5$eV experimentally 
\cite{Kino-Fukuyama,Kino-ET}.

From now on,
we analyze the dimer Hubbard model by applying the RG+cRPA method
\cite{RTazai-PRR2021}.
The RG+cRPA method
is an efficient hybrid method between the RG and the RPA
\cite{Tsuchiizu:2013gu,Tsuchiizu:2016ix,Tsuchiizu-CDW,Tazai-FRG}.
Here, we introduce the higher-energy cutoff $\Lambda_0 \ (=2)$.
The RG flow will stop for 
$\Lambda_l\lesssim \w_c$ with $\w_c={\rm max}\{ T,\gamma\}$,
where $\gamma \ (\propto|{\rm Im}\Sigma|)$ is the quasi-particle damping rate.
Considering large $\gamma$ in $\kappa$-(BEDT-TTF)$_2$X,
we introduce the low-energy cutoff
$\w_c=\pi T$ in the RG equation of the four-point vertex $\Gamma$
in calculating Fig. \ref{fig:RG} (a)
by following Refs. \onlinecite{Tsuchiizu:2016ix,Tsuchiizu-CDW}.

First, we calculate
various spin and charge susceptibilities 
with the form factor $f_{\k,\q}$; $\chi^{c(s)}_f(\q)$
introduced in Eq. (\ref{eqn:chif}).
By analyzing the following from factors
$f=1$, $\sqrt{2}\sin k_x$, $\sqrt{2}\sin k_y$, 
$\cos k_x - \cos k_y$ and $2\sin k_x \sin k_y$,
we find that 
the conventional spin susceptibility $\chi^{s}(\q)$ 
($=\chi_f^{s}(\q)$ with $f=1$)
and the $d$-wave bond susceptibility $\chi^{\rm BO}(\q)$
($=\chi_f^{c}(\q)$ with $f=\cos k_x - \cos k_y$) 
strongly develop.
Other susceptibilities remain small in the present study.

\begin{figure}[htb]
\includegraphics[width=8cm]{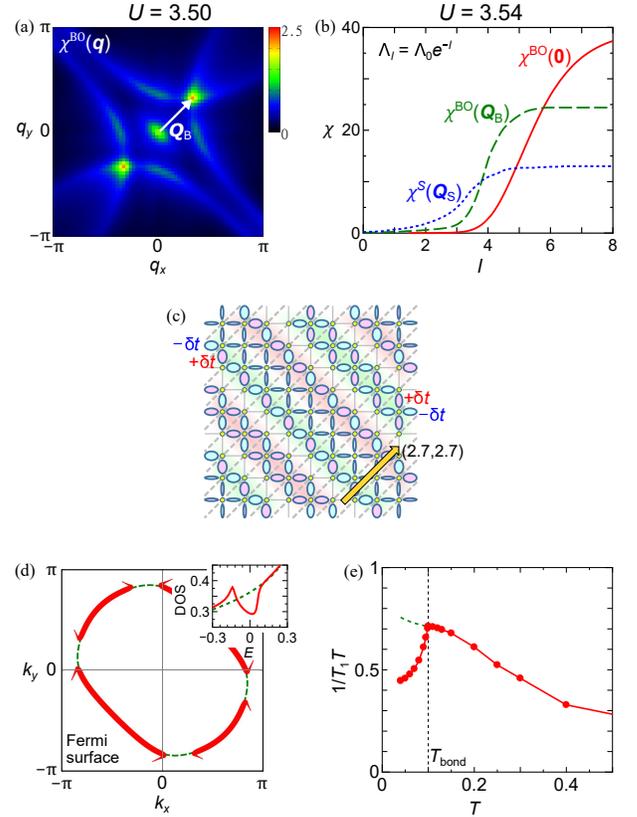}
\caption{
(a) $\q$-dependences of $\chi^{\rm BO}(\q)$
obtained by the RG+cRPA method at $U=3.5$.
(b) The RG flow for spin and BO susceptibilities at $U=3.54$.
(c) Schematic BO pattern at $\q=\Q_{\rm B}$ in real space,
where $\bm{\lambda}\approx(8/3,8/3)$ is the wavelength vector.
(d) Obtained Fermi arc structure in the unfolded zone.
The pseudogap in the DoS with $f^{\rm max}=0.1$ is shown in the inset.
(e) Obtained $1/T_1T$, where $T^*$ is the BO transition temperature.
}
\label{fig-BEDT2}
\end{figure}

In Fig. \ref{fig-BEDT2} (a), we plot $\q$-dependences of 
$\chi^{\rm BO}(\q)$ at $U=3.5$.
We reveal the development of 
$\chi^{\rm BO}(\q)$ at $\q=\Q_{\rm B}\approx (3\pi/8,3\pi/8)$
in addition to $\q=(0,0)$.
The obtained strong bond fluctuations originate from the VCs
that are dropped in the RPA.
(We note that 
$\chi^{\rm BO}(\q)$ is not enhanced at all in the RPA and FLEX.)

The $\chi^{\rm BO}(\q)$ strongly develops by increasing $U$.
Figure \ref{fig-BEDT2} (b) shows the RG flow of the 
susceptibilities in the case of $U=3.54$.
In this case, the bond susceptibility exceeds the spin one
after completing the renormalization.
We see that $\chi^{s}(\Q_{\rm S})$ starts to increase
in the early stage of the renormalization,
by reflecting the major nesting of the FS at $\q=\Q_{\rm S}$.
Next, $\chi^{\rm BO}(\Q_{\rm B})$ starts to increase for $l\gtrsim3$,
and it exceeds $\chi^{s}(\Q_{\rm S})$ at $l\sim4$.
Finally, 
$\chi^{\rm BO}(\bm{0})$
starts to increase for $l\gtrsim 4$ ($\Lambda_l\lesssim 0.037$),
because the renormalization of any $\q=\bm{0}$ susceptibility
occurs only for $\Lambda_l\lesssim T$.
All susceptibilities saturate for $l\gtrsim8$ 
($\Lambda_l\lesssim 0.7\times10^{-3}$).
The final result in Fig. \ref{fig-BEDT2} (a)
is given at $l\approx9$.
Thus, all $\chi^{s}(\Q_{\rm S})$, $\chi^{\rm BO}(\Q_{\rm B})$ and 
$\chi^{\rm BO}(\bm{0})$ strongly develop at $U=3.54$.

Figure \ref{fig-BEDT2} (c)
shows the schematic $d$-wave BO pattern at $\q=\Q_{\rm B}$.
Here, each red (blue) ellipse represents the increment (decrement)
of the hopping integral $\delta t_{\mu}$ ($\mu=x,y$)
caused by the BO parameters.
The opposite sign between the adjacent $\delta t_{x}$ and $\delta t_{y}$ 
reflects the $d$-wave symmetry of the BO.
The BO parameter causes the pseudogap in the DoS:
Figure \ref{fig-BEDT2} (d) shows the Fermi arc structure
obtained for $f^{\rm max}\equiv \max_\k \{f_{\k,\Q_{\rm B}}\} =0.1$.
Here, the folded band structure under the BO at $\q=\Q_{\rm B}$ 
is ``unfolded'' into the original Brillouin zone \cite{Ku}
to make a comparison with ARPES experiment.
The resultant pseudogap in the DoS is shown in 
the inset of Fig. \ref{fig-BEDT2} (d),
which is consistent with the STM study
\cite{Nomura-STM}.
The BO leads to significant reduction 
of the spin fluctuation strength,
so the $1/T_1T$ will exhibit kink-like pseudogap behavior.
To show that, we calculate the value of $1/T_1T$ in Fig. \ref{fig-BEDT2} (e),
which is defined as
\begin{eqnarray}
\frac{1}{T_1T} \propto \sum_{\q,\a,\b} 
{\rm Im}\left. \chi^s_{\a,\b}(\q,\w)/\w \right|_{\w=0},
\end{eqnarray}
where $\a,\b$ represent the sites in the unit cell under the presence of the BO.
We set $f^{\rm max}=0.2\times{\rm tanh}(1.74\sqrt{(1-T/T^*)}$ 
below the BO transition temperature $T^*=0.1$.
(Here, $2f^{\rm max}(T=0)/T^*=4$.)
The obtained pseudogap behaviors in $1/T_1T$ and DoS 
are consistent with phase-transition-like experimental behaviors
\cite{Kanoda-rev,Kanoda-rev2,Raman}.


Finally, we discuss that the physical origin of the 
large ferro-BO ($\q=\bm{0}$)
and incommensurate-BO ($\q=\Q_{\rm B}$) instabilities
in $\kappa$-(BEDT-TTF)$_2$X model depicted in Fig. \ref{fig-BEDT2}(a).
The obtained large BO instabilities are very similar to those of
cuprate and Fe-based superconductors
given by the RG+cRPA method and the DW equation analysis.
Therefore, the main origin of $\q=\bm{0}$ BO and $\q=\Q_{\rm B}$ BO
would be the paramagnon-interference mechanism.
On the other hand, 
$\chi^{\rm BO}(\bm{0})$ is significantly smaller than 
$\chi^{\rm BO}(\bm{Q}_B)$ in the DW analysis for 
$\kappa$-(BEDT-TTF)$_2$X model \cite{RTazai-PRR2021}.
This result indicates that 
large peak at $\q={\bm0}$ in the present RG study
in Fig. \ref{fig-BEDT2}(a) originates from
the spin and SC fluctuations cooperatively, because
the AL process by SC fluctuations can cause the 
ferro-BO instability as revealed in the previous RG study
 \cite{Tsuchiizu:2013gu}.

\section{TRS broken $p$-wave order: charge loop current}
\label{sec:5}

In previous sections, we explained that exotic 
TRS preserving nonlocal orders
are induced by the quantum interference mechanism.
The obtained nematic ($\q={\bm0}$) and smectic ($\q\ne{\bm0}$) BOs
are widely observed not only in cuprates and iridates,
but also in Fe-based superconductors.
In addition, the sLC naturally explain the pseudogap 
in the DoS in cuprates.

Also, the emergence of TRS breaking charge-current orders
has been actively studied in cuprates and iridates
\cite{TRSB-iridate}.
Especially, the intra-unit-cell cLC order along the nearest Cu-O-O triangles 
\cite{TRSB-neutron1,TRSB-neutron2} 
shown in Fig. \ref{fig1-1} (e) has been actively discussed recently.
In this Varma cLC order,
$p$ orbitals on O atoms contribute to the current order,
so extended Hubbard models with O-site and off-site (Cu-O)
Coulomb interactions given by $U_p$ and $V_{\rm Cu-O}$ may have to be analyzed.
Within the MFA, however,
very huge $V_{\rm Cu-O}$ is required to explain the cLC.
Thus, it is important to find the general mechanism of the cLC order
by going beyond the MFA.

To find a general driving force of the cLC order,
it is useful to study simple theoretical models accurately 
using reliable theoretical method.
Here, we study the quasi-one-dimensional (q1D) Hubbard model
at half-filling ($n=1$) by applying the RG theory,
which becomes more reliable in q1D systems rather than 2D systems.
As a result, we reveal that the spin-fluctuation-driven cLC mechanism,
which is expected to be general in low-dimensional Hubbard models
with geometrical frustrations.

\begin{figure}[htb]
\includegraphics[width=8cm]{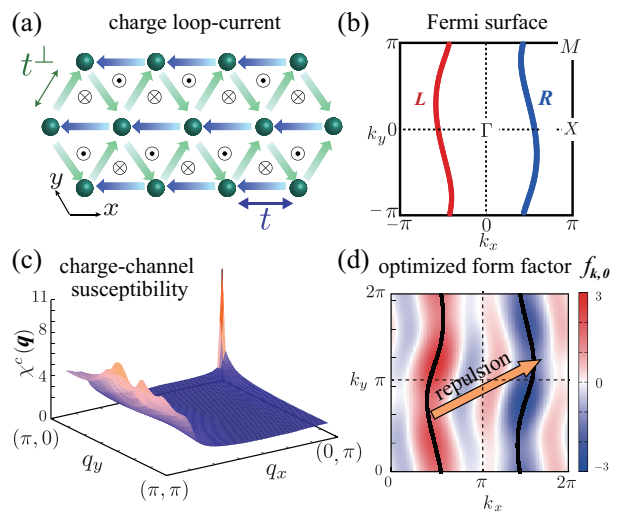}
\caption{(a) Present model and obtained 
charge loop current pattern by RG. 
(b) FS composed of left ($L$) and right ($R$) branches.
(c) Obtained charge-channel susceptibility $\chi^{c}_f(\q)$ with
the form factor. (d) Optimized charge-channel form factor at $\q={\bm0}$,
 which corresponds to the uniform charge-loop current. }
\label{fig:cLC1}
\end{figure}

Below, we study unconventional orders in 
a geometrically frustrated Hubbard model 
\cite{RTazai-PRB2021}.
The Hamiltonian is
$\hat{H}=\hat{H}_{0}+\hat{H}_{I}$,
which is schematically shown in Fig. \ref{fig:cLC1} (a).
The energy dispersion is simply putted by $\xi_{\k}=-2t\cos k_{x}-2t^{\perp}
\{ \cos k_{y}+\cos(k_{x}+k_{y})\}-\mu$ with $t=1$ and the chemical potential
$\mu$.
The inter-chain hopping  $t^{\perp} (\ll 1)$ controls the dimensionality;
$t^{\perp}=0$ corresponds to complete 1D system.
The on-site Coulomb interaction is
$\hat{H}_{I}=\sum_{i}U
n_{i\uparrow}n_{i\downarrow}$ where $i$ is the site index.
The Fermi surface of the present model is shown in Fig. \ref{fig:cLC1} (b)
In the numerical calculation, each
left (L) and right (R) Brillouin zone is divided into $24$ patches.
The logarithmic energy scale for performing the RG is given by
$\Lambda_{l}=\Lambda_0 e^{-l} \,\, (0\leq l \leq l_c )$ for $\Lambda_0=3$ and $\Lambda_{l_c}=T/100$ ($l_c=4.6$).
We consider the half-filling case
and $(t^{\perp},T,U)=(0.2,0.05,2.01)$ is used. 

Based on the RG, we calculate the charge- (spin-) channel susceptibilities with the form factor introduced in Eq. (\ref{eqn:chif}).
The form factor $f_{\k,\q}$ is optimized 
unbiasedly to maximize $\chi^{c(s)}_f(\q)$ at each $\q$-point 
using the Lagrange multipliers method in Sect.\ref{sec:2}.
Figure \ref{fig:cLC1} (c) shows the obtained susceptibility.
The strong charge-channel fluctuations develop at $\q=\bm{0}$,
while the spin fluctuations remain small even
at the nesting vector $\Q_{\rm{S}}=(\pi,\pi/2)$.
Figure \ref{fig:cLC1} (d) shows the $\k$-dependence of the
charge-channel form factor at $\q=\bm{0}$.
For a fixed $k_y$, the obtained result shows $p$-wave symmetry as
\begin{eqnarray}
f_{k_x, k_y,\bm{0}}\simeq -f_{-k_x,k_y,\bm{0}} \hspace{5pt} \propto \sin k_x+b\sin 3k_x .
\label{eq:fkx_cLC} 
\end{eqnarray}
Then, the real-space order parameter is 
$\delta t_{ij}=-\delta t_{ji}$, which leads to
the emergence of ferro-type cLC order. 
Thus, we conclude that the TRS broken $p$-wave cLC phase is strongly stabilized at $(t,t^{\perp})=(0.1,0.2)$.

\begin{figure}[htb]
\includegraphics[width=8cm]{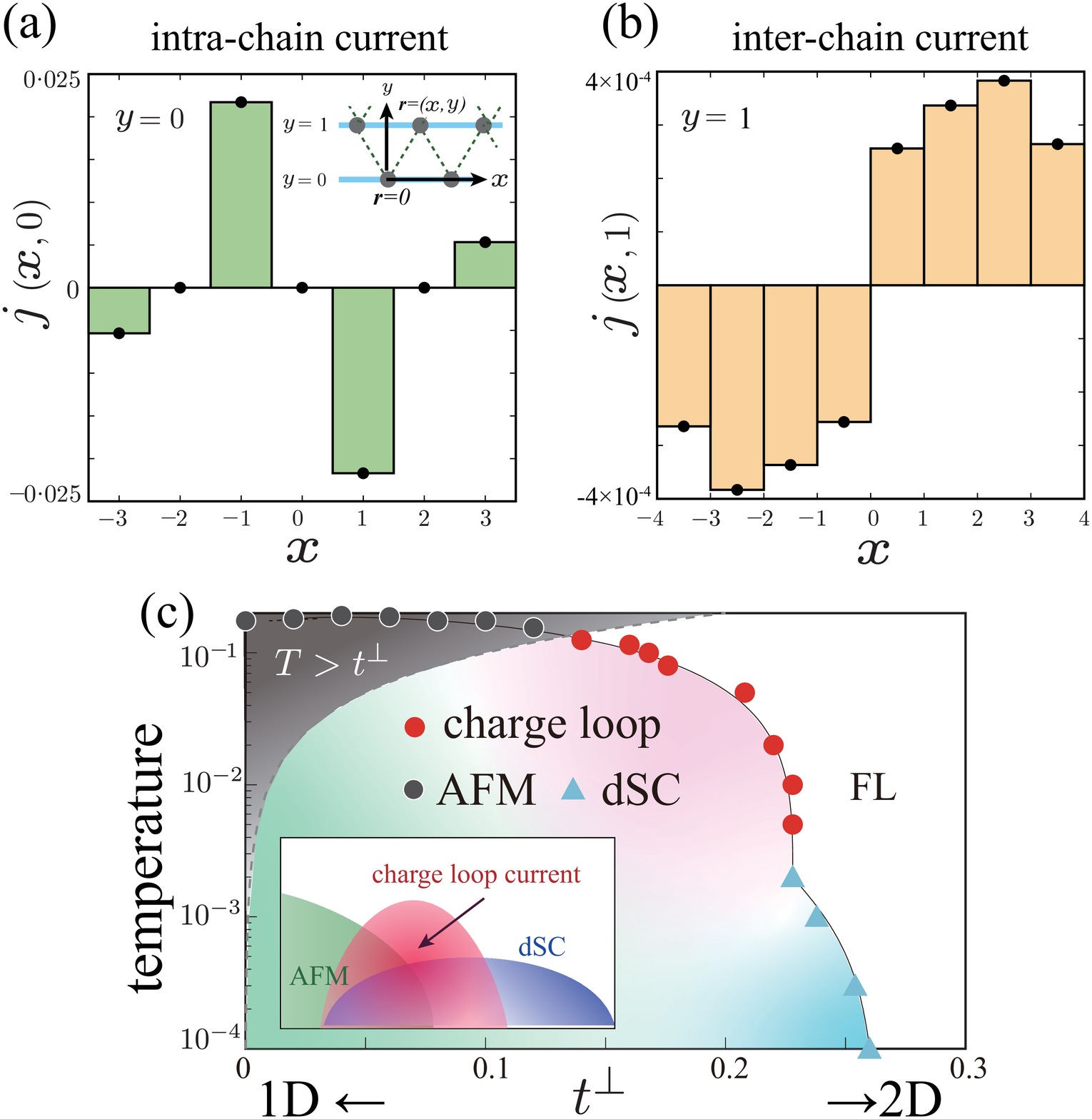}
\caption{(a) Obtained intra-chain current $j(x,0)$ and (b) inter-chain one
$j(x,1)$.
(c) Obtained phase diagram. The charge-loop current order appears
between anti-ferro magnetic and $d$-wave superconducting phases. }
\label{fig:cLC12}
\end{figure}

From the obtained form factor, we calculate the current
from $0$-site  ($\bm{r}=\bm{0}$) to $i$-site ($\bm{r}=\bm{r}_i$) written by
\begin{eqnarray}
j_i= 2i e \left\{ (t_{i0} + \delta t_{i0})G (-\bm{r}_{i})- (t_{0i} + \delta t_{0i})G (\bm{r}_{i}) \right\},
\end{eqnarray} 
where $-e$ is the charge of an electron.
$\delta t_{i0}$ is the Fourier transformation of charge-channel $f_{\k,\q}$
multiplied by the energy scale $\Delta t$.
Note that $\delta t_{i0}$ is pure imaginary
and $\delta t_{i0}=-\delta t_{0i}$ holds.
The equal-time Green function $G (\bm{r}_i)$ in the real space is defined by
\begin{eqnarray}
G (\bm{r}_i)=T\sum_{n,\k} \frac{1}{i\e_n-\xi_{\k}-\Delta t f_{\k,\bm{0}}}
e^{i\k \bm{r}_i}.
\end{eqnarray}
Figures \ref{fig:cLC12} (a) and (b) show the values of 
the intra- and inter-chain current, respectively. 
Here, we put $(e,\Delta t)=(1,0.05)$.
We find that the third-nearest-intra-chain form factor
is significant to obtain the charge-loop current.
In addition, we verified that the macroscopic current is zero due to
the cancellation between intra- and inter-chain current.
Thus, the present result is consistent with Bloch's theory, which
predicts the absence of the macroscopic currents in 
infinite periodic systems \cite{Bohm}.
In Fig. \ref{fig:cLC1} (a), we show the schematic picture of the cLC, which is
a magnetic-octupole-toroidal order. 

Figure \ref{fig:cLC12} (c) is obtained phase diagram in the $T$-$t^{\perp}$ space.
The cLC phase appears around $t^{\perp}\simeq 0.2$
as an intertwined order between antiferro magnetic and $d$-wave superconducting states.
Note that the dark shaded area is 1D Mott insulating phase
that is beyond the scope of the present study \cite{Kishine1,Kishine4}.
As a result, the cLC phase is stabilized in the Fermi liquid region around $t^{\perp} \gg T$.

\begin{figure}[htb]
\includegraphics[width=9cm]{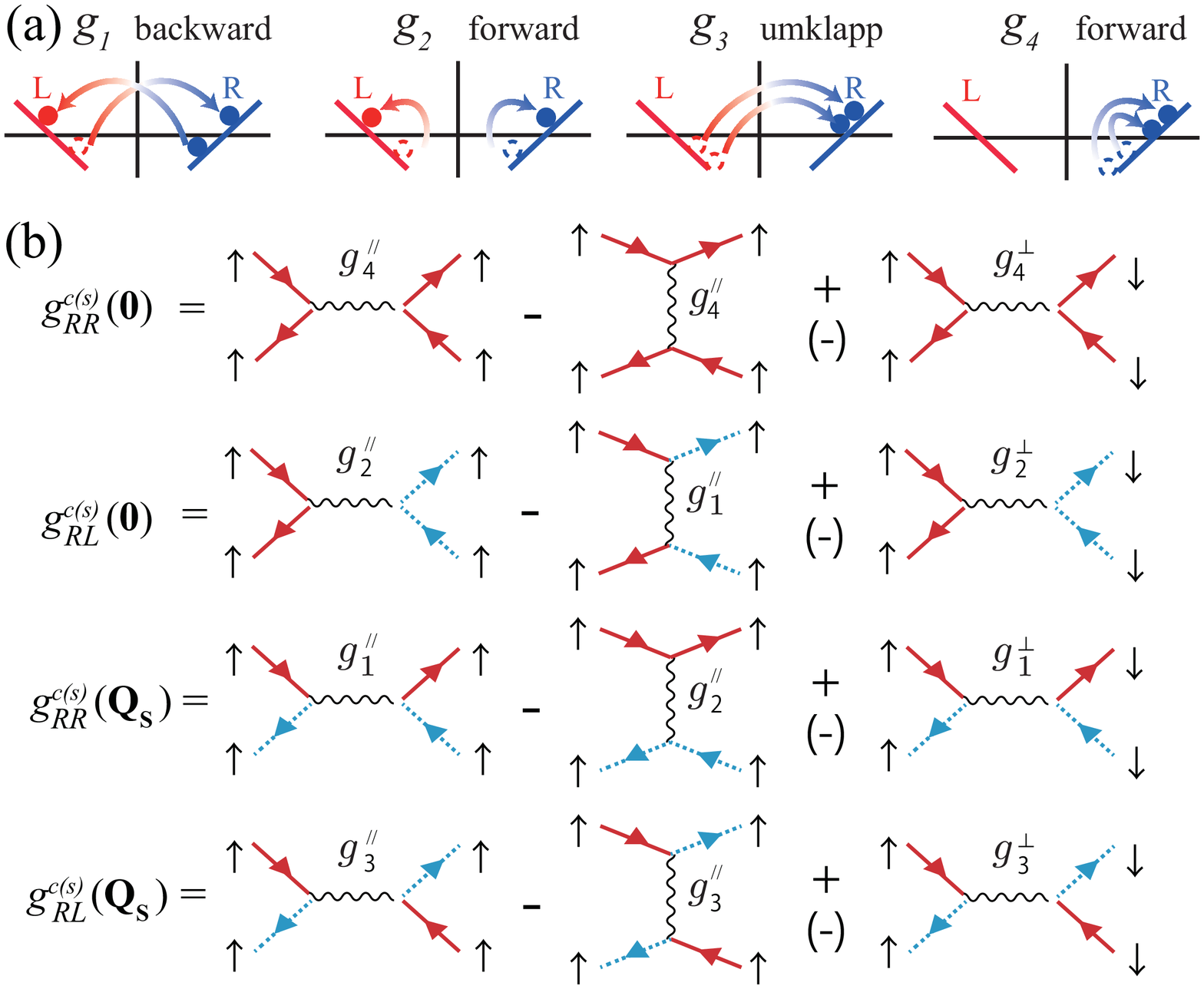}
\caption{(a) Definition of the four-point vertex function $g_{i}$ in the $g$-ology theory.
(b) One-to-one correspondence between $g^{c(s)}_{aa'}(\q)$ and $g_i^{\parallel (\perp)}$. }
\label{fig:cLC2}
\end{figure}

To understand the origin of the cLC, we analyze
the charge- (spin-) channel four-point vertex function based on the $g$-ology theory defined as
\begin{eqnarray}
g^{c(s)}_{a a'}(\q) \equiv \max_{\p\in a,\p' \in a'}\Gamma^{c(s)}_{\p, \p+\q, \p', \p'+\q},
\end{eqnarray}
where $a,a'$ are indices of the branch of the FS and takes $R$ ($p_x>0$) or $L$ ($p_x<0$)
as defined in Fig. \ref{fig:cLC1} (b).
Based on the $g$-ology theory, the $g^{c(s)}$ is
classified into backward ($g_1$), forward ($g_2,g_4$) and umklapp ($g_3$) 
scattering as defined in Fig. \ref{fig:cLC2} (a) \cite{Emery,Bourbonnais,Kishine1,Kishine4,
Suzumura,Suzumura2,SSolyom}.
There is one-to-one correspondence between $g^{c(s)}$ and $g_{i=1-4}$
as diagrammatically shown in Fig. \ref{fig:cLC2} (b), which is described as
\begin{eqnarray}
&\!\!\!\!\!\! g^{c}_{RR}(\bm{0})\approx 2\pi v_F g_4^{\perp}, \hspace{3pt}
&\!\!\! g^{c}_{LR}(\bm{0})\approx 2\pi v_F(2g_2^{\perp} -g_1^{\perp}) \nonumber \\
&\!\!\!\!\!\! g^{s}_{RR}(\bm{Q}_{\rm{S}})\approx -2\pi v_F g_2^{\perp}, \hspace{3pt}
&\!\!\! g^{s}_{LR}(\bm{Q}_{\rm{S}})\approx  -2\pi v_F g_3^{\perp},
\label{eq:golo1}
\end{eqnarray}
where $g^{\perp (\parallel)}$ stands for the four-point vertex function with parallel (anti-parallel) spin.
To derive Eq. (\ref{eq:golo1}), we use the SU(2)-symmetry and anti-commutation relation of the fermion, which leads to 
\begin{eqnarray}
g_{1}^{\perp}-g_{2}^{\perp}=g_1^{\parallel}-g_2^{\parallel}\label{eq:golosu2}.
\end{eqnarray}
This relation is equivalent to that of Eq. (\ref{eqn:SU2}).
Note that the $g^{\parallel}_{3}$ and $g^{\parallel}_{4}$ do not affect the 
physical quantity due to the anti-commutation relation.
Thus, all physical quantities are written by using $g^{\perp}$ without $g^{\parallel}$.

\begin{figure}[htb]
\includegraphics[width=9cm]{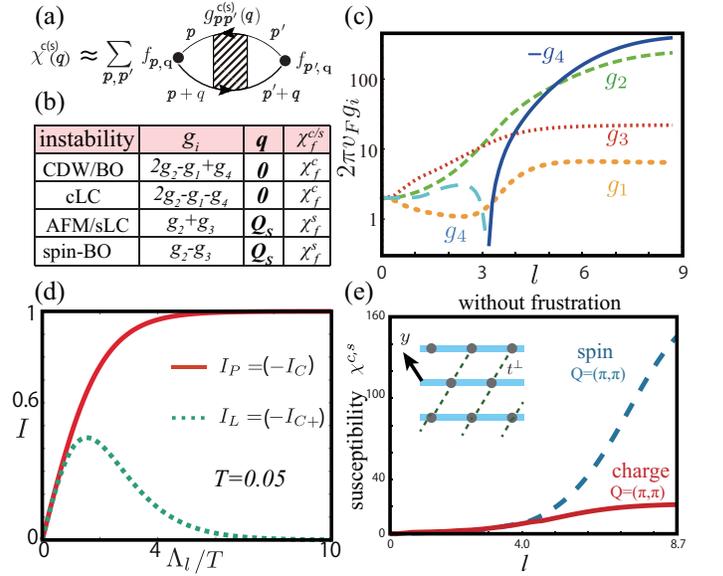}
\caption{(a) Diagrammatic expression of susceptibility with form factor.  The shaded box is the four-point vertex function by RG. (b) Classification of the charge- and spin- channel instabilities. (c) Obtained RG flow of $g_i^{\perp}$.
(d) $\Lambda_l$-dependence of the 
$I_{P}=-I_{C}$ (red solid line) 
and $I_{L}=-I_{C+}$ (green dotted line).
(e) Maximum values of $\chi^{c(s)}_{f} (\q)$ without the geometrical frustration. }
\label{fig:cLC3}
\end{figure}

Both $\chi^{c}_f(\bm{0})$ and $\chi^{s}_f(\bm{Q}_{\rm{S}})$ 
are derived from the RG equations (\ref{eq:chi-RG})-(\ref{eq:R0}).
For a qualitative analysis,
we introduce the following simplified expressions:
\begin{eqnarray}
\hspace{-10pt}\chi^{c}_f(\bm{0})&\sim& -\left\{
(f_{R,\bm{0}})^2 g^{c}_{R R}(\bm{0})+f_{L,\bm{0}} f_{R,\bm{0}} g^{c}_{L R}(\bm{0})\right\} 
\nonumber \\
&&\times (W_{R,R}^{+}(l^*))^2, 
\label{eq:chic-simple} \\
\hspace{-10pt}\chi^{s}_f(\bm{Q}_{\rm{S}})&\sim&  -\left\{
(f_{R,\bm{0}})^2 g^{s}_{R R}(\bm{Q}_{\rm{S}})+f_{L,\bm{0}} f_{R,\bm{0}} g^{s}_{L R}(\bm{Q}_{\rm{S}})  \right\} 
\nonumber \\
&&\times (W_{R,L}^{+}(l^*))^2, 
\label{eq:chis-simple}
\end{eqnarray}
where $l^*$ is an appropriate scaling parameter 
with $T\ll \Lambda_{l^*} \ll E_F$, and
\begin{eqnarray}
W_{\p,\p'}^{\pm}(l)= T\sum_{\bm{k} \bm{k}' n}G(\bm{\k},\e_n)G(\bm{\k}',\pm\e_{n})
\Omega_{\bm{p}}(\bm{k})\Omega_{\bm{p}'}(\bm{k}'),
\end{eqnarray}
where $G(\bm{k},\e_n)= (i\e_n-\xi_{\bm{k}})^{-1}
\theta(\Lambda_{l}-|\xi_{\k}|)$,
and $\Omega_{\bm{p}}(\bm{k})= 1 \ (0)$ only if the momentum $\bm{k}$ 
is inside (outside) of the $\bm{p}$-patch.
Figure \ref{fig:cLC3} (a) represents 
the diagrammatic expression of Eqs. (\ref{eq:chic-simple})
and (\ref{eq:chis-simple}).

In the odd-parity case for $f_{R,\bm{0}}=-f_{L,\bm{0}}$,
the cLC susceptibility is derived from 
${\chi}^{c}_f(\bm{0})$ as
\begin{eqnarray}
{\chi}^{\rm cLC}(\bm{0})\propto
(f_{R,\bm{0}})^2 (-g_{4}^{\perp} +2g_{2}^{\perp}-g_1^{\perp}).
\end{eqnarray}
Thus, the uniform cLC phase appears due to the
coupling constant $-g_4^{\perp}+2g_2^{\perp}-g_1^{\perp}$.
On the other hand, in the even-parity case for $f_{R,\bm{0}}=f_{L,\bm{0}}$,
the AFM susceptibility is derived from ${\chi}^{s}_f(\bm{Q}_{\rm{S}})$ as
\begin{eqnarray}
{\chi}^{\rm AFM}(\bm{Q}_{\rm{S}}) \propto (f_{R,\bm{0}})^2 (g_2^{\perp}+g_3^{\perp}).
\end{eqnarray}
Thus, AFM susceptibility is enlarged due to $g_2^{\perp}+g_3^{\perp}$.
The classification of general instabilities for $\q=0,\Q_{\rm S}$ are summarized in Fig. \ref{fig:cLC3} (b).

Figure \ref{fig:cLC3} (c) shows the obtained RG flow of $g_i^{\perp}$.
We find that $g_4^{\perp}$ ($g_2^{\perp}$) has large negative (positive) value 
as $l$ increases, while $g_1^{\perp}$ and $g_3^{\perp}$
are quite small. Thus, the strong enhancement of the cLC susceptibility originates from $g_2^{\perp}$ and $g_4^{\perp}$. 

\begin{figure*}[htb]
\includegraphics[width=17cm]{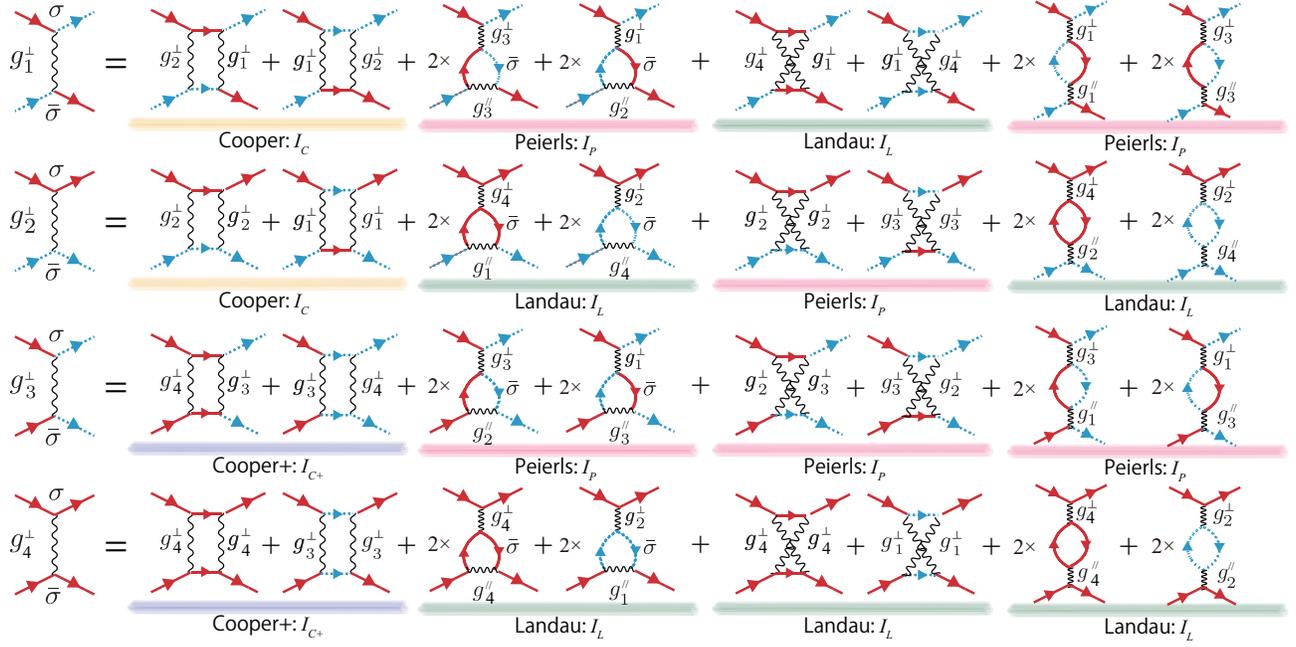}
\caption{One-loop RG equation for the four-point vertex function $g^{\perp}_{i}$ in $g$-ology theory.}
\label{fig:cLC5}
\end{figure*}

To understand the large positive (negative) value of $g_2^{\perp} \ (g_4^{\perp})$, we consider the classical one-loop RG equation for $g_i^{\perp}$, which is given as
\begin{eqnarray}
\frac{dg_{1}^{\perp}}{dl}&=&2g_1^{\perp} g_2^{\perp} I_{C}+2g_1^{\perp}( g_2^{\parallel}-g_1^{\parallel}) I_{P}+2g_1^{\perp} g_4^{\perp} I_{L}, \nonumber \\ 
\frac{dg_2^{\perp}}{dl}&=&(g_2^{\perp}g_2^{\perp} +g_1^{\perp}g_1^{\perp}) I_{C} +2 g_4^{\perp}(g_1^{\parallel}-g_2^{\parallel}) I_{L}\nonumber \\ &+&
(g_2^{\perp} g_2^{\perp}  +g_3^{\perp} g_3^{\perp} ) I_{P}, \nonumber  \\
\frac{dg_3^{\perp}}{dl}&=&2g_3^{\perp} g_4^{\perp} I_{C+}+2g_3^{\perp}(g_2^{\parallel}-
g_1^{\parallel})I_{P}+2g_3^{\perp}g_2^{\perp}I_{P},\nonumber 
\\
\frac{dg_4^{\perp}}{dl}&=&(g_4^{\perp} g_4^{\perp}+g_3^{\perp}g_3^{\perp}) I_{C+}\nonumber \\ &+&
2g_2^{\perp}(g_1^{\parallel}-g_2^{\parallel})I_{L}+(g_4^{\perp} g_4^{\perp}+g_1^{\perp}g_1^{\perp}) I_{L}. \label{eq:goloperp}
\end{eqnarray}
Their diagrammatic expressions are given in Fig. \ref{fig:cLC5}.
Here, 
$I_P$ and $I_L$ denotes the Peierls and Landau channel terms
due to the particle-hope loop diagram, and
$I_{C}$ and $I_{C+}$ are the Cooper and Cooper+ channel ones
\cite{Emery,Bourbonnais,Kishine1,Kishine4,Suzumura,Suzumura2,SSolyom}.
They are expressed as
\begin{eqnarray}
I_{P(L)} &\equiv &2\pi v_F \cdot dW_{\bm{R},\bm{L}(\bm{R})}^{+}/dl, \\
I_{C(C+)}& \equiv &2\pi v_F \cdot dW_{\bm{R},\bm{L}(\bm{R})}^{-}/dl.
\end{eqnarray}

In the present cLC mechanism, we find that Cooper channel scatterings 
are negligible since the same cLC phase is obtained even if we
neglect $I_{C}$ and $I_{C+}$.
Thus, the RG equation becomes simpler as
\begin{eqnarray}
\frac{dg_2}{dl}&=&(g^2_2 +g^2_3) I_{P} +(-2g_2 g_4+2g_1g_4) I_{L},
\label{eq:goloperp2-2} \\
\frac{dg_4}{dl}&=& (g^2_4 +g^2_1-2g^2_2 +2g_1g_2) I_{L},
\label{eq:goloperp2}
\end{eqnarray}
where $g_i$ stands for $g^{\perp}_i$ for the simplicity and 
the SU(2) condition is used.
Then, $g_2$ is enhanced by the Peierls-channel term $(g^2_2 +g^2_3) I_{P}$, while it is suppressed by $-2g_2 g_4 I_{L}$ \cite{Bourbonnais}.
On the other hand, $g_4$ reaches the negative value due to the Landau-channel term $-2g^2_2 I_{L}$.
In the 1D region, it is well known that only $g_2$ is dominant.
Therefore, the enhancement of the Landau channel scattering is a
key fact to obtain the cLC phase.

Moreover, the Landau-channel scattering becomes important
in the presence of geometrical frustration.
In fact, finite $t^{\perp}$ violate the perfect nesting condition, and therefore
$g_2$  is relatively suppressed than the 1D system at $\Lambda_l < t^{\perp}$ \cite{Emery,Bourbonnais,Kishine1,Kishine4,Suzumura,Suzumura2}.
Then, Landau-channel scattering can be enlarged at low-energies ($\Lambda_l < T$) without prohibited by the SDW.
Figure \ref{fig:cLC3} (d) exhibits the
$\Lambda_{l}$-dependence of $I_{L(P)}$ in the linear dispersion model,
which is given by
\begin{eqnarray}
I_{P}&=&\tanh(\Lambda_l/2T), \\
I_{L}&=&(\Lambda_l/2T)\cosh^{-2}(\Lambda_l/2T).
\end{eqnarray}
Thus, the Landau-channel scattering 
becomes as important as the Cooper- (Peierls-) scattering 
in the lower energy region
\cite{fuseya_g4}.
To verify the importance of geometrical frustration,
we calculate the $\chi^{c(s)}_{f}$ by dropping the frustration
($t^{\perp}=0$) in Fig. \ref{fig:cLC3} (e).
In this case, only spin susceptibility develops, 
while the charge one is quite small.
Thus, the cLC order can overcome the SDW order 
in the presence of geometrical frustration.

Next, we verify the spin-fluctuation-driven cLC mechanism
based on the Fermi liquid theory valid for two-dimensional systems.
The important roles of spin fluctuations 
on the $d$-wave superconductivity and the
non-Fermi liquid-type behaviors 
(such as $\rho\propto T$ and $R_H\propto T^{-1}$)
have been discovered previously
\cite{Kontani-rev,Moriya,Yamada,Sato-RH}.
Here, in order to clarify the important role of spin fluctuations
on the cLC formation, we solve the DW equation for
the charge-channel form factor \cite{RTazai-PRB2021}:
\begin{eqnarray}
\lambda_{\q}  f_{\k,\q}=\sum_{\k'} f_{\k',\q} L_{\k',\q} \left(-\frac{3}{2}V^{s}_{\k-\k'} -\frac{1}{2}V^{c}_{\k-\k'} \right),\label{eq:cdw}
\end{eqnarray} 
where $\lambda_{\q}$ is the eigenvalue
and $L_{\k,\q}\equiv (n_{\k_{-}}-n_{\k_{+}})/(\xi_{\k_{+}}- \xi_{\k_{-}}) >0$ with Fermi distribution function $n_{\k}$.
The interaction $V^{c(s)}_{\q}\equiv -(+)U+U^2 \chi^{c(s)}(\q)$ is calculated by the RPA,
We find that the
cLC solution $f_{\k,\q}\propto \sin k_x$ at $\q={\bm 0}$
gives the largest eigenvalue
due to large $V^{s}_{\k-\k'}$ at $\k-\k'\approx\pm\Q_{\rm S}$
 \cite{RTazai-PRB2021}.
In solving the DW equation,
higher-order fluctuation-exchange processes
with respect to $\chi^s(\bm{Q}_{\rm S})$ are generated.
The even (odd)-order processes
gives the inter-branch repulsion $g_2>0$ 
(intra-branch attraction $g_4<0$),
consistently with the $g$-ology analysis
in Fig. \ref{fig:cLC3} (c).

In conclusion, we proposed the microscopic origin of the 
cLC phase based on the RG theory with
optimized form factor.
By virtue of this method, the ferro-type cLC order is obtained with high accuracy
in a simple frustrated chain Hubbard model.
Especially, the geometrical frustration helps the strong enhancement of the 
forward scattering ($g_2$ and $g_4$) via Landau channel scattering.
The present study indicates that the cLC can emerge in various metals near
the magnetic quantum criticality with geometrical frustration.
Thus, the present proposed mechanism can be essential origin of the cLC phases reported in such as cuprates, iridates, and their related materials.

\section{Summary}
\label{sec:6}
Exotic symmetry breaking phenomena,
such as the nematic/smectic BOs and charge/spin current orders,
have been recently reported in many strongly correlated metals.
In this article, we discuss the variety of exotic orderings
in terms of the symmetry breaking in self-energy $\delta t_{i,j}^{c,s}$
($i\ne j$) in a unified way.
(Its Fourier transformation gives the form factor $f_{\k,\q}^{c,s}$.)
Since these exotic order cannot be explained within the
mean-field-level approximations, 
we analyzed beyond-mean-field electron correlations
by applying the RG theory.

Based on the RG theory, 
we found that various types of exotic orders originate from
the quantum interference shown in Fig. \ref{fig-interference}.
Due to this mechanism,
nematic ($\q={\bm0}$) and smectic ($\q\ne{\bm0}$) 
bond orders with $d$-wave form factor 
$f_{\k,\q} \propto\cos k_x-\cos k_y$
appear in both cuprates and $\kappa$-(BEDT-TTF)$_2$X.
The derived bond order naturally explains the 
pseudogap behaviors in these compounds.
The quantum interference mechanism also causes
the sLC order, which naturally explains the pseudogap in the DoS
in cuprates.
Recently, emergence of the three-dimensional CDW phase under the magnetic field
\cite{cuprate-3D,cuprate-3D2} and uniaxial stress \cite{cuprate-3D3}
has been reported by resonant x-ray measurements in cuprates. 
It is an important future problem to explain these experiments by using the present fRG method.

In addition, we discussed the emergence of TRS-breaking 
charge-current orders.
The emergence of exotic orders 
has been discussed by performing precise RG analysis 
of the q1D Hubbard model.
We revealed the spin-fluctuation-driven charge loop-current mechanism,
which is expected to be general in low-dimensional Hubbard models
with geometrical frustrations.
Thus, 
rich quantum phase transition with $d$- and $p$-wave form factors
are driven by the paramagnon interference 
in cuprates and their related materials.

Finally, we comment that the
quantum interference (Fig. \ref{fig-interference})
is significant in $f$-electron systems with strong spin-orbit interaction.
Based on this mechanism, the
multipole-fluctuation-pairing mechanism 
has been discussed in Refs. \onlinecite{RTazai-CeCu2Si2-1,RTazai-CeCu2Si2-2},
and the fully-gapped $s$-wave superconductivity without sign reversal
in CeCu$_2$Si$_2$ is satisfactorily explained.
Also, quadrupole and hexadecapole ordering in CeB$_6$ is also studied 
\cite{RTazai-CeB6}.


\acknowledgements
We are grateful to S. Onari for useful discussions.
This work is supported by Grants-in-Aid for Scientific Research (KAKENHI)
Research (No. JP20K22328, No. JP20K03858, No. JP19H05825, No. JP18H01175, JP16K05442)
from MEXT of Japan.


\end{document}